\documentclass[a4paper,5p,twocolumn,number,sort&compress,fleqn]{elsarticle}
\usepackage{multirow,amsmath}
\pdfoutput=1

\usepackage[sups,scaled=0.96]{XCharter}
\usepackage[charter,expert]{mathdesign}
\usepackage[labelfont=bf]{caption}
\usepackage{hyperref}
\hypersetup{
pdfauthor={Jindrich Kolorenc, http://orcid.org/0000-0003-2627-8302},
pdftitle={Theory of resonant x-ray emission spectra in compounds with
  localized f electrons},
pdfkeywords={RXES} {RIXS} {dynamical mean-field theory} {Anderson
  impurity model} {intermediate valency} {praseodymium},
pdfstartview={FitH}
}

\def\rmd{\mathrm{d}}
\def\rmi{\mathrm{i}}

\def\O{\mathop{\mathrm{O}}\nolimits}
\def\0{\phantom{0}}

\usepackage{microtype}
\makeatletter
\def\ps@pprintTitle{%
     \let\@oddhead\@empty
     \let\@evenhead\@empty
     \def\@oddfoot{\footnotesize\itshape
       Preprint submitted to \ifx\@journal\@empty Elsevier
       \else\@journal\fi\hfill\@date}
     \let\@evenfoot\@oddfoot}

\AtEndDocument{%
 \immediate\write\@auxout{\string\citation{apsrev41Control}}%
}%
\makeatother

\journal{Physica B}
\date{June 30, 2017}

\begin{document}

\begin{frontmatter}

\title{Theory of resonant x-ray emission spectra in compounds with localized
f~electrons}
\author{Jind\v rich Koloren\v c}
\address{Institute of Physics, The Czech Academy of Sciences, Na Slovance
  2, 182 21 Praha, Czech Republic}

\begin{abstract}
I discuss a theoretical description of the resonant x-ray emission
spectroscopy (RXES) that is based on the Anderson impurity model. The
parameters entering the model are determined from material-specific
LDA+DMFT calculations. The theory is applicable across the whole f
series, not only in the limits of nearly empty (La, Ce) or nearly full
(Yb) valence f~shell. Its performance is illustrated on the
pressure-enhanced intermediate valency of elemental praseodymium. The
obtained results are compared to the usual interpretation of RXES,
which assumes that the spectrum is a superposition of several signals,
each corresponding to one configuration of the 4f~shell. The present
theory simplifies to such superposition only if nearly all effects of
hybridization of the 4f~shell with the surrounding states are
neglected. Although the assumption of negligible hybridization sounds
reasonable for lanthanides, the explicit calculations show that it
substantially distorts the analysis of the RXES data.
\end{abstract}

\begin{keyword}
RXES \sep RIXS \sep dynamical mean-field theory \sep Anderson impurity
model \sep intermediate valency \sep praseodymium
\end{keyword}

\end{frontmatter}


\section{Valence histogram and its measurements by core-level
  spectroscopy}
\label{sec:intro}

The 4f electrons in solids are typically localized and behave almost
as if they were in a free atom. In some cases, the position of the 4f
level relative to the other electronic states in a crystal is such
that two configurations, say 4f${}^m$ and 4f${}^{m+1}$, are nearly
degenerate. Then the 4f shell is in a mixed state referred to as the
intermediate valency \cite{lawrence1981}. A related mechanism that
induces an admixture of several configurations in a single 4f shell is
hybridization with the other electronic states. Although this
hybridization is often essentially negligible at ambient conditions,
it can be enhanced by lattice compression that eventually leads to
delocalization of the 4f states \cite{mcmahan2009}. In general, the
hybridization mixes more than two 4f${}^m$ configurations, each
contributing with some weight $w_m$. The width of the histogram
defined by these weights, the so-called \emph{valence histogram}, then
reflects the strength of the hybridization in a given compound.

\begin{figure}
\centering
\includegraphics[width=0.9\linewidth]{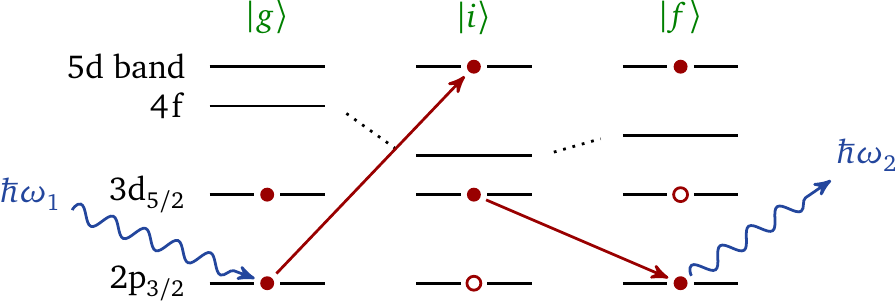}
\caption{\label{fig:RXES1p}A cartoon of the RXES process in a
lanthanide atom: an absorption of a photon~$\hbar\omega_1$ to the
ground state~$|g\rangle$ is followed by an emission of
a~photon~$\hbar\omega_2$ from the intermediate state~$|i\rangle$. Due
to the Coulomb attraction of the core hole, the 4f level is pulled
down in energy by $U_\text{2p4f}$ in the intermediate
state~$|i\rangle$ and by $U_\text{3d4f}$ in the final
state~$|f\rangle$.}
\end{figure}

\begin{figure}
\centering
\includegraphics[width=0.9\linewidth]{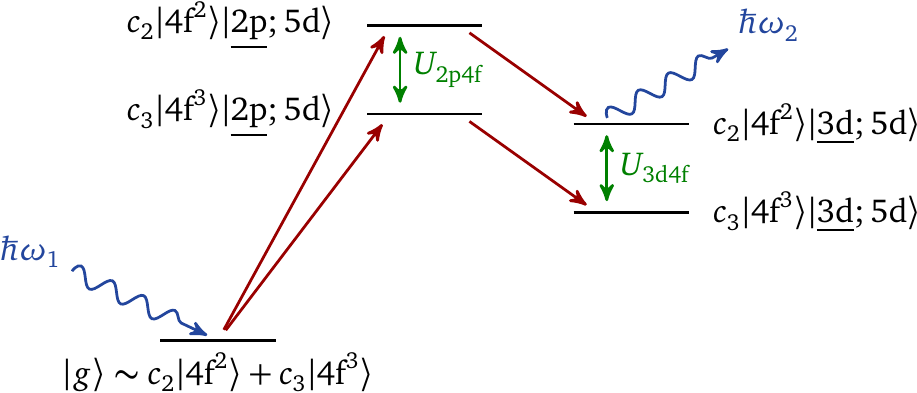}
\caption{\label{fig:RXEStotE}Total-energy diagram corresponding to
  Fig.~\ref{fig:RXES1p}. When the ground state of the 4f
  shell is a linear combination of two configurations (the
  amplitudes~$c_m$ are given as $\sqrt{w_m}$), the intermediate and
  final many-electron states split due to the core-valence Coulomb
  interactions. Underlining indicates a~hole in the particular
  single-electron level.}
\end{figure}

The resonant x-ray emission spectroscopy (RXES, also known as the
core-to-core resonant inelastic x-ray scattering, RIXS) is a direct
probe of unoccupied electronic states. A~measurement, in which the
energy $\hbar\omega_1$ of the incident photons is scanned across the L
edge, maps the 5d density of states. See Fig.~\ref{fig:RXES1p}: an
electron from the 2p${}_{3/2}$ core level is photoexcited to an empty
5d${}_{5/2}$ state (L${}_3$~edge) and this excitation is followed by
relaxation processes that fill the 2p${}_{3/2}$ core hole. In the
particular setup shown in Fig.~\ref{fig:RXES1p}, a detector is tuned
to monitor the photoemission at the L${}_\alpha$~line
($\text{2p}_{3/2}\to\text{3d}_{5/2}$). Although the 4f states are not
directly involved in the excitations, they do influence the measured
spectra. The localized 4f electrons are attracted to the localized
core hole, which modifies the total energy. This modification is
different for different 4f${}^m$~configurations and hence the
absorption edge splits when more configurations are mixed in the
initial state (Fig.~\ref{fig:RXEStotE}). The measured RXES spectra
thus yield an information about the valence histogram
\cite{kotani2001,rueff2010}.

The pressure evolution of the valence histogram in elemental
lanthanides was theoretically investigated with the aid of
material-specific dynamical-mean-field theory (DMFT)
\cite{mcmahan2009}. Subsequently, RXES measurements at the L edge of
praseodymium were performed \cite{bradley2012}, yielding a very good
agreement with the theory. Using a theoretical method very similar to
\cite{mcmahan2009} it was argued that elemental americium approaches
intermediate-valence regime when compressed \cite{savrasov2006} but
L-edge RXES experiments did not detect any sign of this effect
\cite{heathman2010}, possibly due to a limited resolution. Since then,
more L-edge RXES measurements of actinide compounds were made
\cite{booth2014,booth2016} where extraction of the valence histogram
was possible. Some of the conclusions of these actinide measurements
were challenged by RXES data recently collected at the M edges
\cite{kvashnina2017}. Indeed, the interpretation of the spectra in
\cite{booth2014,booth2016} (as well as in \cite{bradley2012}) assumed
a very weak hybridization of the valence f shell with the surrounding
states, which is not likely to be very accurate in actinides.

In this paper, I formulate a theory of RXES that does not make the
assumption of weak hybridization. It generalizes the method previously
used for interpreting RXES measurements on cerium and its compounds
\cite{gunnarsson1983,nakazawa1996,rueff2006} to localized f shells
with more electrons. In addition, I extract a large part of the
parameters entering this theory from first-principles calculations. I
apply this approach to the elemental praseodymium under pressure where
both theory \cite{mcmahan2009} and experiment \cite{bradley2012} are
available for comparison.


\section{Electronic structure of praseodymium under pressure}
\label{sec:Pr}

Praseodymium crystallizes in the dhcp structure at ambient
conditions. When compressed at room temperature, the crystal structure
successively changes to fcc at about 5~GPa, to distorted fcc at
10~GPa, and to orthorombic at 20~GPa. The transition at 20~GPa is
accompanied by an approximately 12\% volume collapse
\cite{baer2003,chesnut2000}.


All calculations presented in this paper are performed for the fcc
lattice (space group Fm$\bar{3}$m). Since the difference between the
fcc and distorted fcc phases is very small (the $c/a$ ratio in the
distorted structure is only 2\% or less larger than in the ideal
close-packed lattice \cite{baer2003}), this setup covers the pressure
range from 5 to 20 GPa. The lattice constants corresponding to several
pressures are listed in Tab.~\ref{tab:GeomWeights}.

\subsection{Valence-band electronic structure: LDA+DMFT}
\label{sec:DMFT}

\begin{table}[b]
\caption{\label{tab:GeomWeights}Lattice constants~$a$ and lattice
volumes~$V$ corresponding to four investigated pressures. The data are
obtained from the Murnaghan equation of state fitted to the
room-temperature measurements combined from \cite{baer2003} and
\cite{chesnut2000}. Also shown are LDA+DMFT results for the average
filling of the praseodymium 4f shell $n_\text{f}$ together with
weights $w_1$, $w_2$ and $w_3$ of 4f${}^1$, 4f${}^2$ and 4f${}^3$
configurations.}
\begin{tabular*}{\columnwidth}{@{\quad}@{\extracolsep{\stretch{1}}}{l}@{\extracolsep{\stretch{1}}}*{4}{c}@{\quad}}
\hline
$P$ (GPa)        & 7.5   & 13    & 21    & 24 \\
$V$/atom (\AA${}^3$)  & 27.37 & 24.23 & 20.98 & 20.02 \\
$a$ (\AA)        & 4.784 & 4.594 & 4.378 & 4.311 \\
$n_\text{f}$ 
      & 2.06\0 & 2.07\0 & 2.09\0 & 2.09\0 \\
$w_1$ & 0.009  & 0.015  & 0.022  & 0.023  \\
$w_2$ & 0.918  & 0.903  & 0.871  & 0.865  \\
$w_3$ & 0.072  & 0.082  & 0.105 &  0.109  \\
\hline
\end{tabular*}
\end{table}

I start the investigation with all-electron calculations of the band
structure in the local-density approximation (LDA) \cite{perdew1992}
taking into account scalar-relativistic effects as well as the
spin-orbital coupling. I employ the \emph{WIEN2k} package
\cite{wien2k} with the following parameters: the radius of the
muffin-tin spheres is $R_\text{MT}=2.7\, a_\text{B}$, the basis-set
cutoff $K_\text{max}$ is defined with $R_\text{MT}\times
K_\text{max}=10.5$, and the Brillouin zone is sampled with 8000 k
points (256 k points in the irreducible wedge). The LDA bands of the
6s, 4f and 5d character are subsequently mapped onto a tight-binding
model with the aid of the \emph{Wannier90} code
\cite{mostofi2008,kunes2010} (34 bands were included in the
disentanglement procedure \cite{souza2001}).

The resulting model $\hat H_\text{LDA}$ then serves as the basis for
LDA+DMFT calculations. The LDA mean-field terms that correspond to the
Coulomb interaction among the 4f~electrons are replaced with an
explicit two-body interaction vertex $\hat U$. This replacement yields
a multiband Hubbard model $\hat H_\text{Hub}=\hat
H_\text{LDA}+\sum_n\bigl(\hat U^n-U^n_\text{DC}\bigr)$ where $n$ runs
over all praseodymium atoms and $\hat U_\text{DC}$, the so-called
double-counting correction, approximates the mean-field terms to be
removed \cite{lichtenstein1998}. The local Coulomb repulsion takes the
form
\begin{equation}
\label{eq:vertex}
\hat U^n= \frac{1}{2}
\sum_{\substack {m m' m''\\  m''' \sigma \sigma'}} 
  U_{m m' m'' m'''} \hat f^{n\dagger}_{m\sigma} \hat f^{n\dagger}_{m' \sigma'}
  \hat f^n_{m'''\sigma'} \hat f^n_{m'' \sigma}\,,
\end{equation}
where $\hat f^n_{m\sigma}$ denotes the f~orbital with magnetic quantum
number $m$ and spin direction $\sigma$ located at lattice site
$n$. The matrix $U_{m m' m'' m'''}$ is assumed to be independent on
pressure. It is parametrized by Slater integrals $F_0=6.0$~eV,
$F_2=10.2$~eV, $F_4=6.3$~eV, $F_6=4.5$~eV. The average Coulomb
parameter $U\equiv F_0$ is set the same as in \cite{mcmahan2009}, the
other values are taken from~\cite{lebegue2006}. The average Hund
exchange corresponding to the given $F_2$, $F_4$ and $F_6$ is
$J=0.82$~eV. The double counting can be expressed as $\hat
U^n_\text{DC}=-U_\text{H} \hat N^n_\text{f}$ where $\hat
N^n_\text{f}=\sum_{m\sigma}\hat f_{m\sigma}^{n\dagger} \hat
f^n_{m\sigma}$ is the number of 4f~electrons at site $n$. The
parameter $U_\text{H}$ is set to $U_\text{H}=7.5$~eV at all
volumes. The rationale for fixing it at a single value comes from the
observation that the computed LDA occupation of the 4f shell
$n_\text{f}$ (projection to the Wannier functions) changes only very
little with compression, from $2.29$ at 30~\AA${}^3$/atom to $2.32$ at
20~\AA${}^3$/atom. The particular value $7.5$ eV was chosen to match
the experimental valence-band photoemission spectra measured at
ambient conditions~\cite{lang1981}.


The Hubbard model $\hat H_\text{Hub}$ is approximately solved using
the dynamical-mean-field theory, that is, the many-body effects
induced by the Coulomb terms $\hat U^n-U^n_\text{DC}$ are taken into
account only locally by means of a site-diagonal
selfenergy~$\hat\Sigma^n(z)$. This selfenergy is computed in an
auxiliary impurity model $\hat H_\text{imp}$ that consists of one
fully interacting f shell (the impurity) at site $i$ embedded in a
self-consistent mean-field medium $\hat H_\text{MF}=\hat H_\text{LDA}
+\sum_{n\not=i}\hat\Sigma^n(z)$ \cite{georges1996}. The impurity model
can be written as $\hat H_\text{imp}=\hat H_\text{imp}^{(0)}+\hat
U-\hat U_\text{DC}$ where $\hat H_\text{imp}^{(0)}$ is a
non-interacting part, and the Coulomb terms $\hat U$ and $\hat
U_\text{DC}$ are the same as defined for the Hubbard model above.

I neglect all non-spherical contributions
(the crystal field) to $\hat H_\text{imp}^{(0)}$, which is
then diagonal in the $|j,j_z\rangle$ basis and reads as
\begin{multline}
\label{eq:Himp0}
\hat H_\text{imp}^{(0)}
=\sum_{j,j_z}\epsilon_j \hat f_{jj_z}^\dagger\hat f_{jj_z}
+\sum_{k=1}^3\sum_{j,j_z}\epsilon_{kj}
  \hat b_{kjj_z}^\dagger\hat b_{kjj_z}\\
+\sum_{k=1}^3\sum_{j,j_z}V_{kj}\bigl(
  \hat f_{jj_z}^\dagger\hat b_{kjj_z}
  + \hat b_{kjj_z}^\dagger\hat f_{jj_z}\bigr)\,.
\end{multline}
The orbitals $\hat f_{jj_z}$ belong to the impurity f~shell, the
orbitals $\hat b_{kjj_z}$ constitute the non-interacting medium (the
so-called bath). Optimally, the model would contain an infinite number
of bath orbitals ($k$ would run to $\infty$) to fully represent $\hat
H_\text{MF}$ in an infinite lattice. The restriction to a finite bath
with $1\le k\le 3$ is a limitation of the exact-diagonalization method
that I employ to solve the model. The comparison to experiment will
provide a justification for the finite bath in the present
application. More thorough discussion of accuracy of the exact
diagonalization in the context of LDA+DMFT can be found elsewhere
\cite{liebsch2012}.

The parameters $\epsilon_j$, $\epsilon_{kj}$ and $V_{kj}$ that enter
Eq.~\eqref{eq:Himp0} are determined by matching the f-shell matrix
elements of the one-particle Green's function computed from $\hat
H_\text{imp}^{(0)}$,
\begin{equation}
G^{(0)}_{\text{f},j}(\rmi\omega_n)=
\biggl(z-\epsilon_j
  -\sum_{k=1}^3\frac{V_{kj}^2}{\rmi\omega_n-\epsilon_{kj}}\biggr)^{-1}\,,
\end{equation}
to the f-shell matrix elements of the so-called bath Green's function
$\mathcal G_j$ that corresponds to the mean-field medium $\hat
H_\text{MF}$ \cite{georges1996}. This matching is implemented as a
minimization of a~cost function in the form
\begin{equation}
\label{eq:fit_cost_func}
\chi_j=\sum_{n=n_\text{min}}^{n_\text{max}}
\biggl|
\frac1{G^{(0)}_{\text{f},j}(\rmi\omega_n)}-
\frac1{\mathcal G_{j}(\rmi\omega_n)}
\biggr|^2
\end{equation}
where $\omega_n$ are Matsubara frequencies \cite{caffarel1994}. I
introduce a lower cutoff $n_\text{min}$ such that the smallest
Matsubara frequency included in the fit is $\omega_n\approx
3$~eV. That way, the high-energy asymptotics of $G^{(0)}_{\text{f},j}$
and $\mathcal G_j$ are as close as possible to each other, which is
advantageous for calculation of thermodynamic quantities such as the
total energy \cite{georges1996}. Although my goal is spectra, the
spectra in question do not involve the 4f~states directly, and it is
the total energy of the 4f shell that plays the important role
(Fig.~\ref{fig:RXEStotE}).


During the DMFT calculations, the selfenergy $\hat\Sigma(z)$ of the
impurity model $\hat H_\text{imp}$ is computed using an in-house
exact-diagonalization code \cite{kolorenc2015b} that combines the
implicitly restarted Lanczos method for finding the many-body spectrum
\cite{arpack} with the band Lanczos method for evaluation of the
one-particle Green's function \cite{meyer1989}. The computational
demands are reduced with the aid of a truncated Hilbert space
\cite{gunnarsson1983,kolorenc2015b}. If the basis states are denoted
as $|f^r b^n\text{\underline{$b$}}^m\rangle$, where $r$ indicates the
number of electrons in the f~states, $n$ the number electrons in the
bath states above the Fermi level, and $m$ the number of holes in the
bath states below the Fermi level, the truncated $N$-electron Hilbert
space is defined as
\begin{equation}
\label{eq:truncHilbert}
\mathcal{H}_N^{(M)}=\bigl\{
|f^{N-N_b^<-n+m}\,b^n\,\text{\underline{$b$}}^m\rangle,\,
0\leq m+n\leq M\bigr\}\,.
\end{equation}
Here $N_b^<$ is the number of bath orbitals below the Fermi level and
the parameter $M$ controls the size of the Hilbert space. All
calculations reported in this paper correspond to $M=2$.


\begin{figure}
\centering
\includegraphics[width=0.9\linewidth]{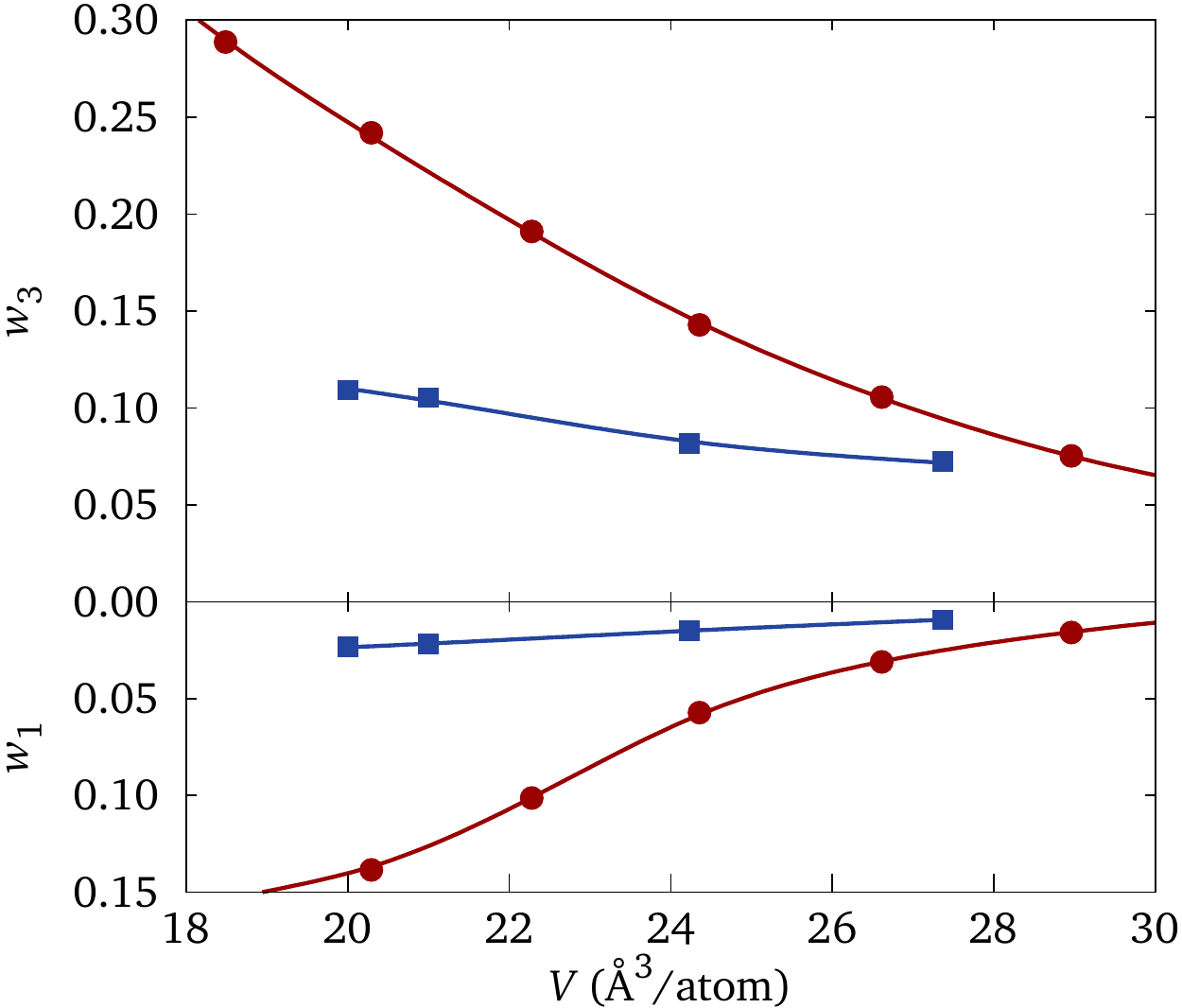}
\caption{\label{fig:weights}The configuration weights $w_1$ and $w_3$
  defined in Eq.~\eqref{eq:weights} as functions of volume. The
  present results (blue squares) are compared to earlier calculations
  (red points) \cite{mcmahan2009}.}
\end{figure}

The results of the LDA+DMFT calculations are summarized in
Fig.~\ref{fig:weights} and in Tab.~\ref{tab:GeomWeights}. The table
lists the pressure evolution of the valence histogram
\begin{equation}
\label{eq:weights}
w_m=\langle\psi_0|\delta(\hat N_\text{f}-m)|\psi_0\rangle\,,\quad
m=0,1,2,\dots
\end{equation}
that comprises weights~$w_m$ of the individual 4f${}^m$ configurations
that contribute to the ground state~$|\psi_0\rangle$ of the impurity
model~$\hat H_\text{imp}$. The average filling of the 4f shell, also
included in Tab.~\ref{tab:GeomWeights}, can then be written as
$n_\text{f}=\langle\psi_0|\hat N_\text{f}|\psi_0\rangle=\sum_m m w_m$.
Figure~\ref{fig:weights} shows the pressure dependence of $w_1$ and
$w_3$ that measure the departure from the 4f${}^2$ atomic limit. The
present results are compared to the earlier LDA+DMFT calculations
\cite{mcmahan2009}. Apparently, there is a large discrepancy, which I
will return to in Sec.~\ref{sec:discussion}.


\subsection{Resonant x-ray emission spectra}
\label{sec:RXES}

The resonant x-ray emission spectroscopy can be simulated in an impurity
model
\begin{multline}
\hat H=\hat H_\text{imp}
+\epsilon_\text{2p}\hat p^{\dagger}\hat p
+\epsilon_\text{3d}\hat d^{\dagger}\hat d
+\sum_k\epsilon_k\hat d_k^{\dagger}\hat d_k\\
+\Bigl[U_\text{2p4f}\bigl(\hat p^{\dagger}\hat p-1\bigr)
+U_\text{3d4f}\bigl(\hat d^{\dagger}\hat d-1\bigr)\Bigr]\hat N_\text{f}
\end{multline}
that represents an extension of the impurity model $\hat H_\text{imp}$
that appeared in Sec.~\ref{sec:DMFT} as a part of the LDA+DMFT
theory. The model $\hat H_\text{imp}$ is supplemented with the
following orbitals located at the impurity site: the 2p${}_{3/2}$ core
state $\hat p$, the 3d${}_{5/2}$ core state $\hat d$, and the
5d${}_{5/2}$ band $\hat d_k$. The core electrons and the 4f~electrons
are localized, and they repel each other by the Coulomb interaction
parametrized by $U_\text{2p4f}$ for the 2p${}_{3/2}$ state and by
$U_\text{3d4f}$ for the 3d${}_{5/2}$ state. The degeneracy of the core
levels is neglected and hence there is only one core-valence Slater
integral. The interaction between the 4f~electrons and 5d~electrons
can be neglected thanks to the delocalized character of the 5d states.

The cross section of the RXES process is approximated by the
Kramers--Heisenberg formula~\cite{sakurai_aqm,kotani2001}
\begin{multline}
\label{eq:KramersHeisenberg}
\sigma(\omega_1,\omega_2) = \sum_f\biggl|\sum_i\frac{%
\langle f|\hat T_2|i\rangle\langle i|\hat T_1|g\rangle}{%
E_i-E_g-\omega_1-\rmi\Gamma_\text{2p}}\biggr|^2
\\ \times
\frac{\Gamma_\text{3d}/\pi}{(\omega_1-\omega_2-E_f+E_g)^2+\Gamma_\text{3d}^2}\,,
\end{multline}
where $|g\rangle$ is the initial ground state with completely filled
core states and empty 5d states, $|i\rangle$ is an intermediate state
with a 2p${}_{3/2}$ core hole and an excited electron in the 5d band,
and $|f\rangle$ is a final state with a 3d${}_{5/2}$ core hole and one
5d electron. The total energies of these states are denoted as $E_g$,
$E_i$ and~$E_f$; $\Gamma_\text{2p}$ and $\Gamma_\text{3d}$ are half
widths at half maximum of the 2p${}_{3/2}$ and 3d${}_{5/2}$ hole
states, and $\omega_1$ and $\omega_2$ stand for energies of the
incident and emitted photons. Finally, $\langle f|\hat T_2|i\rangle$
and $\langle i|\hat T_1|g\rangle$ are the dipole matrix elements. If I
consider only the angular part of the dipole operator, I have $\hat
T_1=\sum_k\hat d_k^{\dagger}\hat p$ and $\hat T_2=\hat p^{\dagger}\hat
d$.

The initial state can be written as $|g\rangle=\hat p^{\dagger}\hat
d^{\dagger}|\psi_0\rangle$ where $|\psi_0\rangle$ is the correlated
ground state of the impurity model $\hat H_\text{imp}$, $\hat
H_\text{imp}|\psi_0\rangle=E_0|\psi_0\rangle$. Apparently, $|g\rangle$
is an eigenstate of the full hamiltonian~$\hat H$, that is, $\hat
H|g\rangle=E_g|g\rangle$ with
$E_g=E_0+\epsilon_\text{2p}+\epsilon_\text{3d}$. The
Kramers--Heisenberg formula as written in
Eq.~\eqref{eq:KramersHeisenberg} is useful when it is possible to
explicitly calculate all intermediate and final states as eigenstates
of the hamiltonian $\hat H$. This is not the case in the present
application since the Hilbert space is too large. For the purposes of
the Krylov-subspace method, which I intend to use, I rewrite
Eq.~\eqref{eq:KramersHeisenberg} into an operator form.

First, I can replace the sums $\sum_i$ and $\sum_f$ with sums
over all eigenstates of $\hat H$ since the transition operators $\hat T_1$
and $\hat T_2$ will select the right states anyway,
\begin{align}
\label{eq:KHwithT}
\sigma(\omega_1,\omega_2) &=
\sum_k\langle\psi_0|\hat d
\hat d_k
\frac1{\hat H-E_g-\omega_1+\rmi\Gamma_\text{2p}}
\hat d^{\dagger}\hat p \nonumber\\
&\qquad\times
\frac{\Gamma_\text{3d}/\pi}{(\omega_1-\omega_2-\hat H+E_g)^2+\Gamma_\text{3d}^2}
\nonumber\\
&\qquad\times
\hat p^{\dagger}\hat d
\frac1{\hat H-E_g-\omega_1-\rmi\Gamma_\text{2p}}
\hat d_k^{\dagger}
\hat d^{\dagger}|\psi_0\rangle\,.
\end{align}
The fermionic operators associated with the uncorrelated states can be
commuted toward $\langle\psi_0|$ where they all annihilate. Along the
way, the full impurity hamiltonian $\hat H$ is replaced by two smaller
hamiltonians,
\begin{align}
\sigma(\omega_1,\omega_2)=&
\sum_k\langle\psi_0|
\frac1{\hat H_\text{imp}'-E_0-\epsilon_\text{2p}
  +\epsilon_k-\omega_1+\rmi\Gamma_\text{2p}}
\nonumber\\
&\times
\frac{\Gamma_\text{3d}/\pi}{(\omega_1-\omega_2-\hat H_\text{imp}''+E_0+\epsilon_\text{3d}-\epsilon_k)^2+\Gamma_\text{3d}^2}
\nonumber\\
&\times
\frac1{\hat H_\text{imp}'-E_0-\epsilon_\text{2p}
  +\epsilon_k-\omega_1-\rmi\Gamma_\text{2p}}
|\psi_0\rangle\,,
\label{eq:threeG}
\end{align}
where $\hat H_\text{imp}'=\hat H_\text{imp}-U_\text{2p4f}\hat N_\text{f}$
and $\hat H_\text{imp}''=\hat H_\text{imp}-U_\text{3d4f}\hat N_\text{f}$.
These hamiltonians correspond to the intermediate and final states
where the 4f level is pulled down by $U_\text{2p4f}$ and
$U_\text{3d4f}$ due to interaction with the core hole
(Fig.~\ref{fig:RXES1p}).


The cross section $\sigma(\omega_1,\omega_2)$ takes the form of a
diagonal matrix element of a product of three functions of two
non-commuting operators, which is a rather complex object. If,
however, the two hamiltonians $\hat H_\text{imp}'$ and $\hat
H_\text{imp}''$ were the same, that is, if
$U_\text{2p4f}=U_\text{3d4f}=U_\text{cv}$, the formula
would reduce to an expectation value of a function of a
single operator, $\langle\psi_0|F(\hat
H_\text{imp}')|\psi_0\rangle$. Such matrix element is considerably
simpler and can be evaluated with the aid of the Krylov-subspace
approximation, see \ref{app:Krylov}. The assumption of equal
core-valence Coulomb interactions $U_\text{cv}$ in the intermediate
and final states is well justified in the case of the deep 2p and
3d core states. In cerium, $U_\text{2p4f}$ and $U_\text{3d4f}$
differ only by about $0.5$~eV \cite{rueff2006}.

Finally, I replace the momentum summation in Eq.~\eqref{eq:threeG}
with an integral over 5d single-particle energies, which yields a
convolution of a many-body expectation value with the density of
5d${}_{5/2}$ states that is taken from the LDA+DMFT calculations,
\begin{multline}
\sigma=
\int\rmd\epsilon g_{\text{5d}_{5/2}}(\epsilon)\langle\psi_0|
\frac1{(\hat H_\text{imp}'-E_0-\epsilon_\text{2p}
  +\epsilon-\omega_1)^2+\Gamma_\text{2p}^2}\\
\times
\frac{\Gamma_\text{3d}/\pi}{(\omega_1-\omega_2-\hat H_\text{imp}'
 +E_0+\epsilon_\text{3d}-\epsilon)^2+\Gamma_\text{3d}^2}
|\psi_0\rangle\,.
\label{eq:RIXSfinal}
\end{multline}

\begin{figure}
\includegraphics[width=0.5\linewidth]{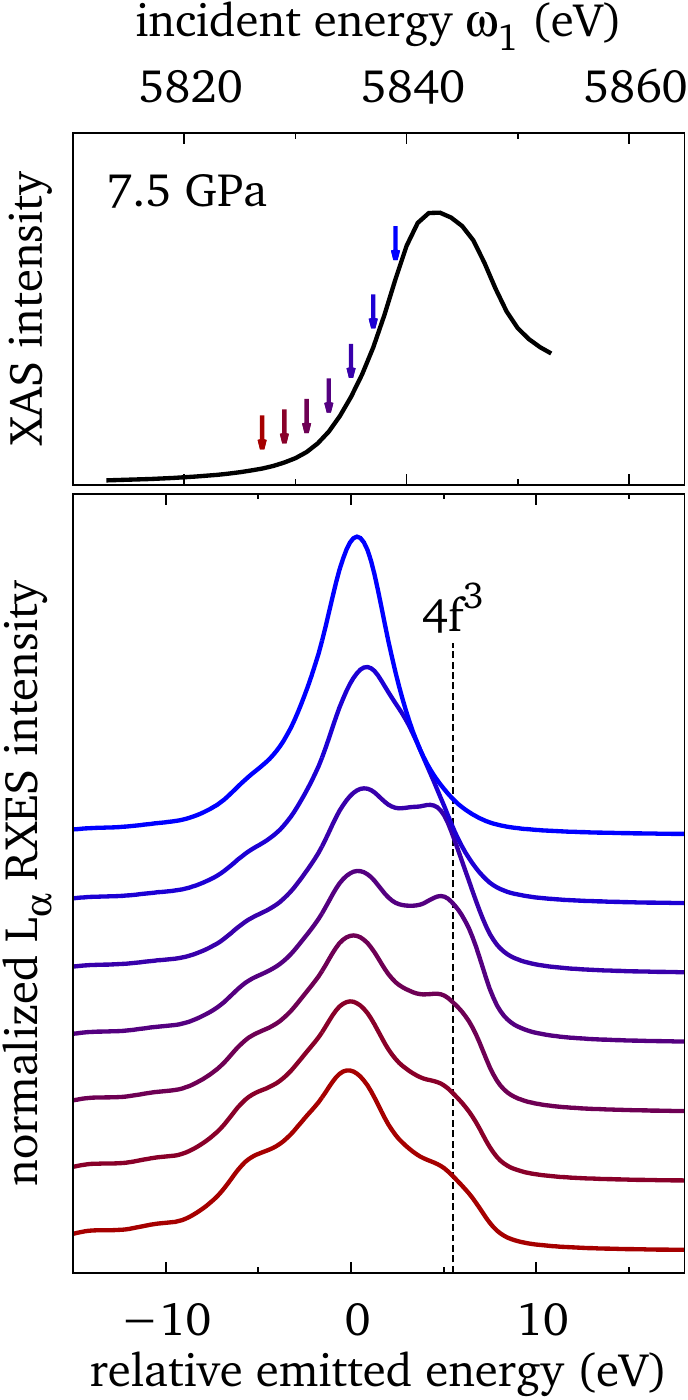}%
\includegraphics[width=0.5\linewidth]{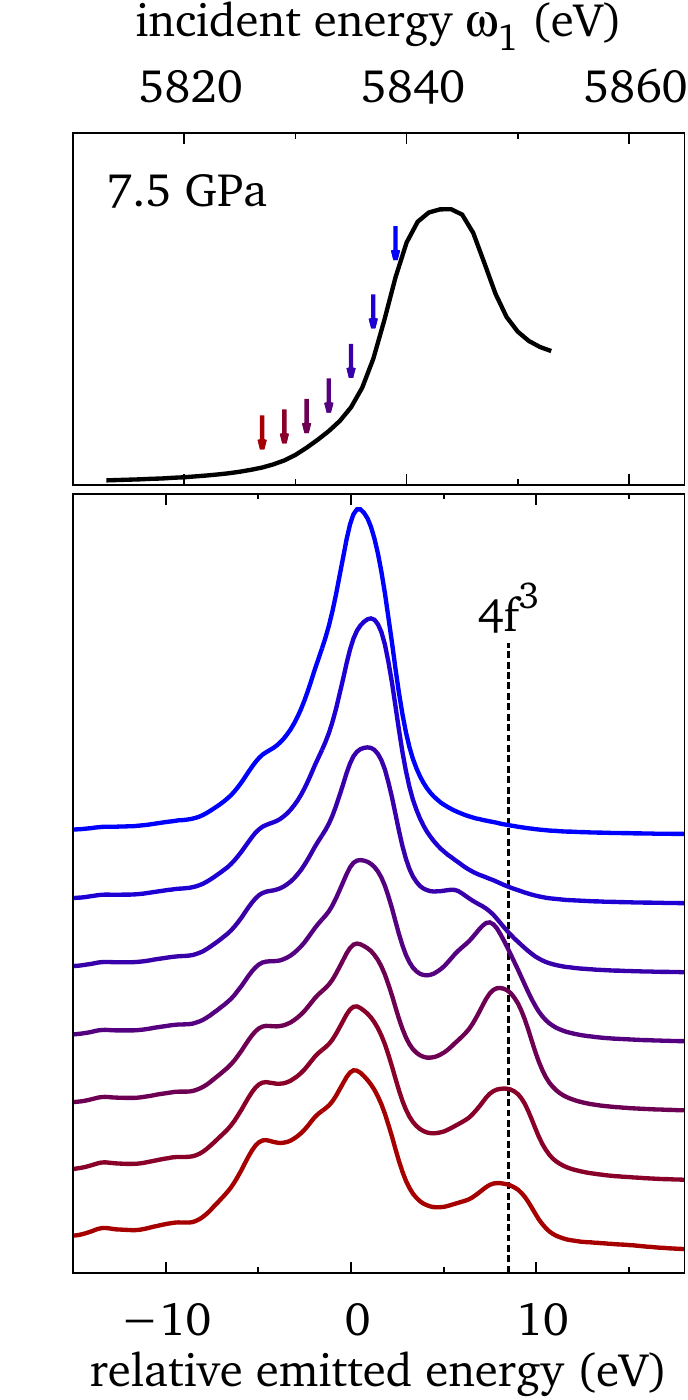}
\caption{\label{fig:scan}Theoretical RXES spectra corresponding to the
  incident energies~$\omega_1$ approaching the L${}_3$ absorption edge
  (top panel). The relative
  emitted energy on the horizontal axis means $\omega_2-\omega_1$
  shifted so that the main peak is centered at zero. \emph{Left}:
  The theory given by
  Eq.~\eqref{eq:RIXSfinal} closely matches the experimental data shown
  in Fig.~1 of~\cite{bradley2012}. \emph{Right}: The simplified theory
  represented by Eqs.~\eqref{eq:RIXSsimple} is significantly worse.}
\end{figure}

\begin{figure}[t]
\centering
\includegraphics[width=0.5\linewidth]{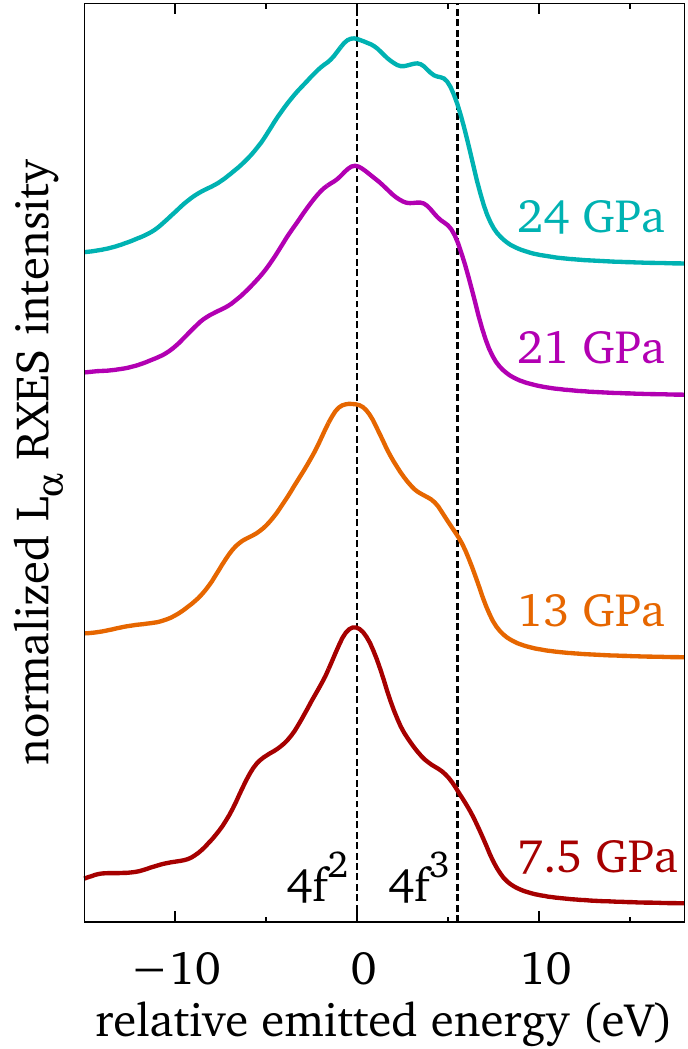}
\caption{\label{fig:pressure}RXES spectrum as a function of
  pressure. The shoulder associated with the 4f${}^3$ component grows
  in a compressed lattice. The theoretical curves accurately reproduce the
  measurements shown in Fig.~2a of~\cite{bradley2012}.}
\end{figure}

The spectra given by Eq.~\eqref{eq:RIXSfinal} are plotted in
Figs.~\ref{fig:scan} and~\ref{fig:pressure}. The parameters not fixed
by the first-principles calculations (Sec.~\ref{sec:DMFT}) are the
core-valence Coulomb parameter, which is assumed to be 25\% larger
than the valence-valence Coulomb parameter, $U_\text{cv}=1.25U=7.5$
eV, and the line widths $\Gamma_\text{2p}=5.0$~eV and
$\Gamma_\text{3d}=1.5$~eV that combine the natural line width and the
instrument resolution ($\Gamma_\text{2p}$ is possibly overestimated
\cite{krause1979}).

Figure~\ref{fig:scan} shows $\sigma(\omega_1,\omega_2)$ at $7.5$~GPa
for incident-photon energy~$\omega_1$ approaching the absorption
edge. One can see how the signal corresponding to the minority
4f${}^3$ configuration is resonantly enhanced before the main edge
corresponding to 4f${}^2$ is reached. Compared to experiment
\cite{bradley2012}, the absolute position of the L${}_3$ edge is
estimated poorly, but the shapes of the spectra for incident energies
at comparable distances from the edge are reproduced rather
accurately. Figure~\ref{fig:pressure} shows RXES spectra at different
pressures, all calculated for $\omega_1=5827$~eV, that is, about 16~eV
below the maximum of the absorption line. Again, the spectra look like
those experimentally observed. Note, however, that the experimental
data shown in Fig.~2a of~\cite{bradley2012} are taken at the incident
energy closer to the maximum of the absorption line.


\section{Discussion}
\label{sec:discussion}

The theoretical approach outlined in Sec.~\ref{sec:Pr} accurately
reproduces the RXES spectra experimentally measured for elemental
praseodymium \cite{bradley2012}. It does so with a 4f valence
histogram \emph{substantially different} from the histogram previously
calculated in \cite{mcmahan2009} as well as from the histogram deduced
from the experimental data in \cite{bradley2012}. The latter two are
compatible to each other, the present theory is at odds with both, see
Tab.~\ref{tab:ratio} for details. My calculations suggest that a
considerably smaller departure from the 4f$^2$ atomic limit than
previously thought is needed to explain the measured RXES.

\begin{table}[b]
\caption{\label{tab:ratio}The weight of the 4f${}^3$
  configuration in the ground state increases with
  pressure. The present theory (DMFT/ED) suggests a noticeably slower
  growth than both the earlier theory (DMFT/QMC) \cite{mcmahan2009}
  and the earlier interpretation of the measured RXES data
  \cite{bradley2012}.}
\begin{tabular*}{\columnwidth}{@{\quad}@{\extracolsep{\stretch{1}}}*{4}{c}@{\quad}}
\hline
\multirow{2}{*}{$P$ (GPa)} & \multicolumn{3}{c}{$w_3/w_2$} \\
\cline{2-4}
 & DMFT/ED & DMFT/QMC \cite{mcmahan2009} & exp. \cite{bradley2012}\\
\hline
24  & 0.126 & 0.42 & 0.46(5) \\
21  & 0.121 & 0.34 & 0.33(5) \\
13  & 0.090 & 0.19 & 0.18(5) \\
7.5 & 0.079 & 0.11 & 0.13(5) \\
\hline
\end{tabular*}
\end{table}

The theory used in \cite{mcmahan2009} is very similar to the method
discussed here in Sec.~\ref{sec:DMFT}. It is also LDA+DMFT, only: (i)
the impurity model is solved by the Hirsch--Fye quantum Monte Carlo
(QMC) instead of the exact diagonalization, (ii) the Hund exchange is
neglected, that is, $F_2=F_4=F_6=0$, and (iii)~the Coulomb parameter
$U$ is volume dependent \cite{mcmahan1998}. The reduction of $U$ due
to more efficient screening in a compressed lattice certainly causes
some increase of $w_1$ and $w_3$, but hardly as much as would be
needed to explain the large discrepancy shown in
Fig.~\ref{fig:weights} and Tab.~\ref{tab:ratio} since $U$ drops only
by 0.5~eV from 30~\AA${}^3$/atom to 20~\AA${}^3$/atom
\cite{mcmahan1998}. A more substantial issue is the calculation of the
weights~$w_m$. Unlike certain variants of the continuous-time QMC
\cite{gull2011}, the Hirsch--Fye QMC does not provide a direct access
to these weights, which were estimated using expressions borrowed from
the atomic limit, thus neglecting hybridization~\cite{mcmahan2009}.

An assumption of vanishingly small hybridization was employed also
when the RXES measurements were analyzed in \cite{bradley2012}: the
spectra were fitted by a linear combination of two signals, and their
intensities were directly interpreted as the weights $w_2$ and
$w_3$. The cross section derived in Sec.~\ref{sec:RXES} displays such
a simple behavior only if: (i)~the hybridization is neglected in $\hat
H_\text{imp}'$, which corresponds to the condition $U_\text{cv}\gg
V_{kj}$, and (ii) the incident energy is far below the absorption
edge, that is, $\epsilon_\text{2p}+\omega_1$ is much larger than
$U_\text{cv}$, $\Gamma_\text{2p}$ and the 5d bandwidth. Under these
conditions, Eq.~\eqref{eq:RIXSfinal} reduces~to
\begin{subequations}
\label{eq:RIXSsimple}
\begin{equation}
\sigma(\omega_1,\omega_2)=\sum_m w_m f(\omega_1-\omega_2+ m U_\text{cv})
\end{equation}
where
\begin{equation}
f(x)=\int\rmd\epsilon\,\frac{g_{\text{5d}_{5/2}}(\epsilon)}{%
(\epsilon_\text{2p}+\omega_1)^2}
\frac1{(x-\epsilon+\epsilon_\text{3d})^2 + \Gamma_\text{3d}^2}\,.
\end{equation}
\end{subequations}
To gauge the consequences of neglecting the hybridization in the
intermediate and final states of RXES, the spectra calculated with the
simplified formula and with Eq.~\eqref{eq:RIXSfinal} are compared in
Fig.~\ref{fig:scan} for otherwise identical parameters. The
differences are sizable.

Both the experiment \cite{bradley2012} and the earlier theory
\cite{mcmahan2009} assume somewhere along the way that the
hybridization of the 4f shell with the surrounding electronic states
is negligibly small, although it is exactly this hybridization that is
the dominant cause of the investigated effect: the nonzero
weights~$w_1$ and~$w_3$. The present theory does not rely on this
assumption.


\section{Summary}
\label{sec:summary}

I have outlined a theoretical model for the resonant x-ray emission
spectra measured at the L edge in f-electron compounds. It builds on
the ideas developed for Ce compounds in the past
\cite{gunnarsson1983,nakazawa1996} and extends them to take into
account the full multiplet structure of the valence f~states. The
theory avoids the assumption of very weak hybridization of the f~shell
with the surrounding states, which is typically made when interpreting
the experimental spectra
\cite{bradley2012,booth2014,booth2016}. Taking compressed praseodymium
as an example, I have illustrated that this assumption can skew the
determination of the valence histogram already in lanthanides, with
the experiments on actinides being affected even more.


\section*{Acknowledgments}
This work was supported by the Czech Science Foundation [grant number
15-05872J]. Access to computing facilities owned by parties and
projects contributing to the National Grid Infrastructure MetaCentrum,
provided under the program Cesnet LM2015042, is appreciated.



\appendix
\section{Krylov-subspace approximation}
\label{app:Krylov}
A diagonal matrix element of a function of an operator~$\hat A$ can be
approximated as
\cite{meyer1989}
\begin{equation}
\label{eq:approxKrylov}
\langle\psi|f(\hat A)|\psi\rangle
= \langle\psi|f(\hat B)|\psi\rangle
+\O\bigl(\hat A^{n+1}\bigr)\,,
\end{equation}
where $\hat B$ is an operator living in a Krylov subspace
\begin{equation}
\mathcal K_n=\bigl\{|\psi\rangle, \hat A|\psi\rangle,
\hat A^2|\psi\rangle,\dots,\hat A^n|\psi\rangle
 \bigr\}\,.
\end{equation}
The operator $\hat B$ is defined as
$B_{ij}=\langle\phi_i|\hat A|\phi_j\rangle$, where $\{|\phi_i\rangle\}$ is
an orthonormal basis of $\mathcal K_n$ such that
$|\phi_1\rangle=|\psi\rangle$. The method is useful if the
approximation is sufficiently accurate already for $n$ so small that
all eigenvalues $b_i$ of $\hat B$ and their corresponding eigenvectors
$|b_i\rangle$ can be explicitly found, and one can use the spectral
representation of Eq.~\eqref{eq:approxKrylov},
\begin{equation}
\langle\psi|f(\hat A)|\psi\rangle
= \sum_{i=1}^n \bigl|\langle\psi|b_i\rangle\bigr|^2 f(b_i)
+\O\bigl(\hat A^{n+1}\bigr)\,.
\end{equation}


\newpage
\def\bibsection{\relax}
\def\bibpreamble{\relax}
\bibliographystyle{apsrev4-1jk}
\bibliography{aim,dft,dmft,lanczos,codes,Pr,rxes,AnO,pu,qm,mixed_valency,am,apsrev41Control}

\begin{thebibliography}{33}%
\makeatletter
\providecommand \@ifxundefined [1]{%
 \@ifx{#1\undefined}
}%
\providecommand \@ifnum [1]{%
 \ifnum #1\expandafter \@firstoftwo
 \else \expandafter \@secondoftwo
 \fi
}%
\providecommand \@ifx [1]{%
 \ifx #1\expandafter \@firstoftwo
 \else \expandafter \@secondoftwo
 \fi
}%
\providecommand \natexlab [1]{#1}%
\providecommand \enquote  [1]{``#1''}%
\providecommand \bibnamefont  [1]{#1}%
\providecommand \bibfnamefont [1]{#1}%
\providecommand \citenamefont [1]{#1}%
\providecommand \href@noop [0]{\@secondoftwo}%
\providecommand \href [0]{\begingroup \@sanitize@url \@href}%
\providecommand \@href[1]{\@@startlink{#1}\@@href}%
\providecommand \@@href[1]{\endgroup#1\@@endlink}%
\providecommand \@sanitize@url [0]{\catcode `\\12\catcode `\$12\catcode
  `\&12\catcode `\#12\catcode `\^12\catcode `\_12\catcode `\%12\relax}%
\providecommand \@@startlink[1]{}%
\providecommand \@@endlink[0]{}%
\providecommand \url  [0]{\begingroup\@sanitize@url \@url }%
\providecommand \@url [1]{\endgroup\@href {#1}{\urlprefix }}%
\providecommand \urlprefix  [0]{URL }%
\providecommand \Eprint [0]{\href }%
\providecommand \doibase [0]{http://dx.doi.org/}%
\providecommand \selectlanguage [0]{\@gobble}%
\providecommand \bibinfo  [0]{\@secondoftwo}%
\providecommand \bibfield  [0]{\@secondoftwo}%
\providecommand \translation [1]{[#1]}%
\providecommand \BibitemOpen [0]{}%
\providecommand \bibitemStop [0]{}%
\providecommand \bibitemNoStop [0]{.\EOS\space}%
\providecommand \EOS [0]{\spacefactor3000\relax}%
\providecommand \BibitemShut  [1]{\csname bibitem#1\endcsname}%
\let\auto@bib@innerbib\@empty
\bibitem [{\citenamefont {Lawrence}\ \emph {et~al.}(1981)\citenamefont
  {Lawrence}, \citenamefont {Riseborough},\ and\ \citenamefont
  {Parks}}]{lawrence1981}%
  \BibitemOpen
  \bibfield  {author} {\bibinfo {author} {\bibfnamefont {J.~M.}\ \bibnamefont
  {Lawrence}}, \bibinfo {author} {\bibfnamefont {P.~S.}\ \bibnamefont
  {Riseborough}},\ and\ \bibinfo {author} {\bibfnamefont {R.~D.}\ \bibnamefont
  {Parks}},\ }\href {\doibase10.1088/0034-4885/44/1/001} {\bibfield  {journal}
  {\bibinfo  {journal} {Rep. Prog. Phys.}\ }\textbf {\bibinfo {volume} {44}},\
  \bibinfo {pages} {1--84} (\bibinfo {year} {1981})}\BibitemShut {NoStop}%
\bibitem [{\citenamefont {McMahan}\ \emph {et~al.}(2009)\citenamefont
  {McMahan}, \citenamefont {Scalettar},\ and\ \citenamefont
  {Jarrell}}]{mcmahan2009}%
  \BibitemOpen
  \bibfield  {author} {\bibinfo {author} {\bibfnamefont {A.~K.}\ \bibnamefont
  {McMahan}}, \bibinfo {author} {\bibfnamefont {R.~T.}\ \bibnamefont
  {Scalettar}},\ and\ \bibinfo {author} {\bibfnamefont {M.}~\bibnamefont
  {Jarrell}},\ }\href {\doibase10.1103/PhysRevB.80.235105} {\bibfield
  {journal} {\bibinfo  {journal} {Phys. Rev. B}\ }\textbf {\bibinfo {volume}
  {80}},\ \bibinfo {pages} {235105} (\bibinfo {year} {2009})},\ \Eprint
  {http://arxiv.org/abs/0909.0539} {arXiv:0909.0539
  [cond-mat.str-el]}\BibitemShut {NoStop}%
\bibitem [{\citenamefont {Kotani}\ and\ \citenamefont
  {Shin}(2001)}]{kotani2001}%
  \BibitemOpen
  \bibfield  {author} {\bibinfo {author} {\bibfnamefont {A.}~\bibnamefont
  {Kotani}}\ and\ \bibinfo {author} {\bibfnamefont {S.}~\bibnamefont {Shin}},\
  }\href {\doibase10.1103/RevModPhys.73.203} {\bibfield  {journal} {\bibinfo
  {journal} {Rev. Mod. Phys.}\ }\textbf {\bibinfo {volume} {73}},\ \bibinfo
  {pages} {203--246} (\bibinfo {year} {2001})}\BibitemShut {NoStop}%
\bibitem [{\citenamefont {Rueff}\ and\ \citenamefont
  {Shukla}(2010)}]{rueff2010}%
  \BibitemOpen
  \bibfield  {author} {\bibinfo {author} {\bibfnamefont {J.-P.}\ \bibnamefont
  {Rueff}}\ and\ \bibinfo {author} {\bibfnamefont {A.}~\bibnamefont {Shukla}},\
  }\href {\doibase10.1103/RevModPhys.82.847} {\bibfield  {journal} {\bibinfo
  {journal} {Rev. Mod. Phys.}\ }\textbf {\bibinfo {volume} {82}},\ \bibinfo
  {pages} {847--896} (\bibinfo {year} {2010})},\ \Eprint
  {http://arxiv.org/abs/0812.0538} {arXiv:0812.0538
  [cond-mat.str-el]}\BibitemShut {NoStop}%
\bibitem [{\citenamefont {Bradley}\ \emph {et~al.}(2012)\citenamefont
  {Bradley}, \citenamefont {Moore}, \citenamefont {Lipp}, \citenamefont
  {Mattern}, \citenamefont {Pacold}, \citenamefont {Seidler}, \citenamefont
  {Chow}, \citenamefont {Rod}, \citenamefont {Xiao},\ and\ \citenamefont
  {Evans}}]{bradley2012}%
  \BibitemOpen
  \bibfield  {author} {\bibinfo {author} {\bibfnamefont {J.~A.}\ \bibnamefont
  {Bradley}}, \bibinfo {author} {\bibfnamefont {K.~T.}\ \bibnamefont {Moore}},
  \bibinfo {author} {\bibfnamefont {M.~J.}\ \bibnamefont {Lipp}}, \bibinfo
  {author} {\bibfnamefont {B.~A.}\ \bibnamefont {Mattern}}, \bibinfo {author}
  {\bibfnamefont {J.~I.}\ \bibnamefont {Pacold}}, \bibinfo {author}
  {\bibfnamefont {G.~T.}\ \bibnamefont {Seidler}}, \bibinfo {author}
  {\bibfnamefont {P.}~\bibnamefont {Chow}}, \bibinfo {author} {\bibfnamefont
  {E.}~\bibnamefont {Rod}}, \bibinfo {author} {\bibfnamefont {Y.}~\bibnamefont
  {Xiao}},\ and\ \bibinfo {author} {\bibfnamefont {W.~J.}\ \bibnamefont
  {Evans}},\ }\href {\doibase10.1103/PhysRevB.85.100102} {\bibfield  {journal}
  {\bibinfo  {journal} {Phys. Rev. B}\ }\textbf {\bibinfo {volume} {85}},\
  \bibinfo {pages} {100102} (\bibinfo {year} {2012})}\BibitemShut {NoStop}%
\bibitem [{\citenamefont {Savrasov}\ \emph {et~al.}(2006)\citenamefont
  {Savrasov}, \citenamefont {Haule},\ and\ \citenamefont
  {Kotliar}}]{savrasov2006}%
  \BibitemOpen
  \bibfield  {author} {\bibinfo {author} {\bibfnamefont {S.~Y.}\ \bibnamefont
  {Savrasov}}, \bibinfo {author} {\bibfnamefont {K.}~\bibnamefont {Haule}},\
  and\ \bibinfo {author} {\bibfnamefont {G.}~\bibnamefont {Kotliar}},\ }\href
  {\doibase10.1103/PhysRevLett.96.036404} {\bibfield  {journal} {\bibinfo
  {journal} {Phys. Rev. Lett.}\ }\textbf {\bibinfo {volume} {96}},\ \bibinfo
  {pages} {036404} (\bibinfo {year} {2006})},\ \Eprint
  {http://arxiv.org/abs/cond-mat/0507552} {arXiv:cond-mat/0507552
  [cond-mat.str-el]}\BibitemShut {NoStop}%
\bibitem [{\citenamefont {Heathman}\ \emph {et~al.}(2010)\citenamefont
  {Heathman}, \citenamefont {Rueff}, \citenamefont {Simonelli}, \citenamefont
  {Denecke}, \citenamefont {Griveau}, \citenamefont {Caciuffo},\ and\
  \citenamefont {Lander}}]{heathman2010}%
  \BibitemOpen
  \bibfield  {author} {\bibinfo {author} {\bibfnamefont {S.}~\bibnamefont
  {Heathman}}, \bibinfo {author} {\bibfnamefont {J.-P.}\ \bibnamefont {Rueff}},
  \bibinfo {author} {\bibfnamefont {L.}~\bibnamefont {Simonelli}}, \bibinfo
  {author} {\bibfnamefont {M.~A.}\ \bibnamefont {Denecke}}, \bibinfo {author}
  {\bibfnamefont {J.-C.}\ \bibnamefont {Griveau}}, \bibinfo {author}
  {\bibfnamefont {R.}~\bibnamefont {Caciuffo}},\ and\ \bibinfo {author}
  {\bibfnamefont {G.~H.}\ \bibnamefont {Lander}},\ }\href
  {\doibase10.1103/PhysRevB.82.201103} {\bibfield  {journal} {\bibinfo
  {journal} {Phys. Rev. B}\ }\textbf {\bibinfo {volume} {82}},\ \bibinfo
  {pages} {201103} (\bibinfo {year} {2010})}\BibitemShut {NoStop}%
\bibitem [{\citenamefont {Booth}\ \emph {et~al.}(2014)\citenamefont {Booth},
  \citenamefont {Medling}, \citenamefont {Jiang}, \citenamefont {Bauer},
  \citenamefont {Tobash}, \citenamefont {Mitchell}, \citenamefont {Veirs},
  \citenamefont {Wall}, \citenamefont {Allen}, \citenamefont {Kas},
  \citenamefont {Sokaras}, \citenamefont {Nordlund},\ and\ \citenamefont
  {Weng}}]{booth2014}%
  \BibitemOpen
  \bibfield  {author} {\bibinfo {author} {\bibfnamefont {C.}~\bibnamefont
  {Booth}}, \bibinfo {author} {\bibfnamefont {S.}~\bibnamefont {Medling}},
  \bibinfo {author} {\bibfnamefont {Y.}~\bibnamefont {Jiang}}, \bibinfo
  {author} {\bibfnamefont {E.}~\bibnamefont {Bauer}}, \bibinfo {author}
  {\bibfnamefont {P.}~\bibnamefont {Tobash}}, \bibinfo {author} {\bibfnamefont
  {J.}~\bibnamefont {Mitchell}}, \bibinfo {author} {\bibfnamefont
  {D.}~\bibnamefont {Veirs}}, \bibinfo {author} {\bibfnamefont
  {M.}~\bibnamefont {Wall}}, \bibinfo {author} {\bibfnamefont {P.}~\bibnamefont
  {Allen}}, \bibinfo {author} {\bibfnamefont {J.}~\bibnamefont {Kas}}, \bibinfo
  {author} {\bibfnamefont {D.}~\bibnamefont {Sokaras}}, \bibinfo {author}
  {\bibfnamefont {D.}~\bibnamefont {Nordlund}},\ and\ \bibinfo {author}
  {\bibfnamefont {T.-C.}\ \bibnamefont {Weng}},\ }\href
  {\doibase10.1016/j.elspec.2014.03.004} {\bibfield  {journal} {\bibinfo
  {journal} {J. Electron. Spectrosc. Relat. Phenom.}\ }\textbf {\bibinfo
  {volume} {194}},\ \bibinfo {pages} {57--65} (\bibinfo {year}
  {2014})}\BibitemShut {NoStop}%
\bibitem [{\citenamefont {Booth}\ \emph {et~al.}(2016)\citenamefont {Booth},
  \citenamefont {Medling}, \citenamefont {Tobin}, \citenamefont {Baumbach},
  \citenamefont {Bauer}, \citenamefont {Sokaras}, \citenamefont {Nordlund},\
  and\ \citenamefont {Weng}}]{booth2016}%
  \BibitemOpen
  \bibfield  {author} {\bibinfo {author} {\bibfnamefont {C.~H.}\ \bibnamefont
  {Booth}}, \bibinfo {author} {\bibfnamefont {S.~A.}\ \bibnamefont {Medling}},
  \bibinfo {author} {\bibfnamefont {J.~G.}\ \bibnamefont {Tobin}}, \bibinfo
  {author} {\bibfnamefont {R.~E.}\ \bibnamefont {Baumbach}}, \bibinfo {author}
  {\bibfnamefont {E.~D.}\ \bibnamefont {Bauer}}, \bibinfo {author}
  {\bibfnamefont {D.}~\bibnamefont {Sokaras}}, \bibinfo {author} {\bibfnamefont
  {D.}~\bibnamefont {Nordlund}},\ and\ \bibinfo {author} {\bibfnamefont
  {T.-C.}\ \bibnamefont {Weng}},\ }\href {\doibase10.1103/PhysRevB.94.045121}
  {\bibfield  {journal} {\bibinfo  {journal} {Phys. Rev. B}\ }\textbf {\bibinfo
  {volume} {94}},\ \bibinfo {pages} {045121} (\bibinfo {year} {2016})},\
  \Eprint {http://arxiv.org/abs/1607.03953} {arXiv:1607.03953
  [cond-mat.str-el]}\BibitemShut {NoStop}%
\bibitem [{\citenamefont {Kvashnina}\ \emph {et~al.}(2017)\citenamefont
  {Kvashnina}, \citenamefont {Walker}, \citenamefont {Magnani}, \citenamefont
  {Lander},\ and\ \citenamefont {Caciuffo}}]{kvashnina2017}%
  \BibitemOpen
  \bibfield  {author} {\bibinfo {author} {\bibfnamefont {K.~O.}\ \bibnamefont
  {Kvashnina}}, \bibinfo {author} {\bibfnamefont {H.~C.}\ \bibnamefont
  {Walker}}, \bibinfo {author} {\bibfnamefont {N.}~\bibnamefont {Magnani}},
  \bibinfo {author} {\bibfnamefont {G.~H.}\ \bibnamefont {Lander}},\ and\
  \bibinfo {author} {\bibfnamefont {R.}~\bibnamefont {Caciuffo}},\ }\href
  {\doibase10.1103/PhysRevB.95.245103} {\bibfield  {journal} {\bibinfo
  {journal} {Phys. Rev. B}\ }\textbf {\bibinfo {volume} {95}},\ \bibinfo
  {pages} {245103} (\bibinfo {year} {2017})},\ \Eprint
  {http://arxiv.org/abs/1706.02920} {arXiv:1706.02920
  [cond-mat.str-el]}\BibitemShut {NoStop}%
\bibitem [{\citenamefont {Gunnarsson}\ and\ \citenamefont
  {Sch\"onhammer}(1983)}]{gunnarsson1983}%
  \BibitemOpen
  \bibfield  {author} {\bibinfo {author} {\bibfnamefont {O.}~\bibnamefont
  {Gunnarsson}}\ and\ \bibinfo {author} {\bibfnamefont {K.}~\bibnamefont
  {Sch\"onhammer}},\ }\href {\doibase10.1103/PhysRevB.28.4315} {\bibfield
  {journal} {\bibinfo  {journal} {Phys. Rev. B}\ }\textbf {\bibinfo {volume}
  {28}},\ \bibinfo {pages} {4315--4341} (\bibinfo {year} {1983})}\BibitemShut
  {NoStop}%
\bibitem [{\citenamefont {Nakazawa}\ \emph {et~al.}(1996)\citenamefont
  {Nakazawa}, \citenamefont {Tanaka}, \citenamefont {Uozumi},\ and\
  \citenamefont {Kotani}}]{nakazawa1996}%
  \BibitemOpen
  \bibfield  {author} {\bibinfo {author} {\bibfnamefont {M.}~\bibnamefont
  {Nakazawa}}, \bibinfo {author} {\bibfnamefont {S.}~\bibnamefont {Tanaka}},
  \bibinfo {author} {\bibfnamefont {T.}~\bibnamefont {Uozumi}},\ and\ \bibinfo
  {author} {\bibfnamefont {A.}~\bibnamefont {Kotani}},\ }\href
  {\doibase10.1143/JPSJ.65.2303} {\bibfield  {journal} {\bibinfo  {journal} {J.
  Phys. Soc. Jpn.}\ }\textbf {\bibinfo {volume} {65}},\ \bibinfo {pages}
  {2303--2310} (\bibinfo {year} {1996})}\BibitemShut {NoStop}%
\bibitem [{\citenamefont {Rueff}\ \emph {et~al.}(2006)\citenamefont {Rueff},
  \citenamefont {Iti\'e}, \citenamefont {Taguchi}, \citenamefont {Hague},
  \citenamefont {Mariot}, \citenamefont {Delaunay}, \citenamefont {Kappler},\
  and\ \citenamefont {Jaouen}}]{rueff2006}%
  \BibitemOpen
  \bibfield  {author} {\bibinfo {author} {\bibfnamefont {J.-P.}\ \bibnamefont
  {Rueff}}, \bibinfo {author} {\bibfnamefont {J.-P.}\ \bibnamefont {Iti\'e}},
  \bibinfo {author} {\bibfnamefont {M.}~\bibnamefont {Taguchi}}, \bibinfo
  {author} {\bibfnamefont {C.~F.}\ \bibnamefont {Hague}}, \bibinfo {author}
  {\bibfnamefont {J.-M.}\ \bibnamefont {Mariot}}, \bibinfo {author}
  {\bibfnamefont {R.}~\bibnamefont {Delaunay}}, \bibinfo {author}
  {\bibfnamefont {J.-P.}\ \bibnamefont {Kappler}},\ and\ \bibinfo {author}
  {\bibfnamefont {N.}~\bibnamefont {Jaouen}},\ }\href
  {\doibase10.1103/PhysRevLett.96.237403} {\bibfield  {journal} {\bibinfo
  {journal} {Phys. Rev. Lett.}\ }\textbf {\bibinfo {volume} {96}},\ \bibinfo
  {pages} {237403} (\bibinfo {year} {2006})},\ \Eprint
  {http://arxiv.org/abs/cond-mat/0602169} {arXiv:cond-mat/0602169
  [cond-mat.str-el]}\BibitemShut {NoStop}%
\bibitem [{\citenamefont {Baer}\ \emph {et~al.}(2003)\citenamefont {Baer},
  \citenamefont {Cynn}, \citenamefont {Iota}, \citenamefont {Yoo},\ and\
  \citenamefont {Shen}}]{baer2003}%
  \BibitemOpen
  \bibfield  {author} {\bibinfo {author} {\bibfnamefont {B.~J.}\ \bibnamefont
  {Baer}}, \bibinfo {author} {\bibfnamefont {H.}~\bibnamefont {Cynn}}, \bibinfo
  {author} {\bibfnamefont {V.}~\bibnamefont {Iota}}, \bibinfo {author}
  {\bibfnamefont {C.-S.}\ \bibnamefont {Yoo}},\ and\ \bibinfo {author}
  {\bibfnamefont {G.}~\bibnamefont {Shen}},\ }\href
  {\doibase10.1103/PhysRevB.67.134115} {\bibfield  {journal} {\bibinfo
  {journal} {Phys. Rev. B}\ }\textbf {\bibinfo {volume} {67}},\ \bibinfo
  {pages} {134115} (\bibinfo {year} {2003})}\BibitemShut {NoStop}%
\bibitem [{\citenamefont {Chesnut}\ and\ \citenamefont
  {Vohra}(2000)}]{chesnut2000}%
  \BibitemOpen
  \bibfield  {author} {\bibinfo {author} {\bibfnamefont {G.~N.}\ \bibnamefont
  {Chesnut}}\ and\ \bibinfo {author} {\bibfnamefont {Y.~K.}\ \bibnamefont
  {Vohra}},\ }\href {\doibase10.1103/PhysRevB.62.2965} {\bibfield  {journal}
  {\bibinfo  {journal} {Phys. Rev. B}\ }\textbf {\bibinfo {volume} {62}},\
  \bibinfo {pages} {2965--2968} (\bibinfo {year} {2000})}\BibitemShut {NoStop}%
\bibitem [{\citenamefont {Perdew}\ and\ \citenamefont
  {Wang}(1992)}]{perdew1992}%
  \BibitemOpen
  \bibfield  {author} {\bibinfo {author} {\bibfnamefont {J.~P.}\ \bibnamefont
  {Perdew}}\ and\ \bibinfo {author} {\bibfnamefont {Y.}~\bibnamefont {Wang}},\
  }\href {\doibase10.1103/PhysRevB.45.13244} {\bibfield  {journal} {\bibinfo
  {journal} {Phys. Rev. B}\ }\textbf {\bibinfo {volume} {45}},\ \bibinfo
  {pages} {13244--13249} (\bibinfo {year} {1992})}\BibitemShut {NoStop}%
\bibitem [{\citenamefont {Blaha}\ \emph {et~al.}(2001)\citenamefont {Blaha},
  \citenamefont {Schwarz}, \citenamefont {Madsen}, \citenamefont {Kvasnicka},\
  and\ \citenamefont {Luitz}}]{wien2k}%
  \BibitemOpen
  \bibfield  {author} {\bibinfo {author} {\bibfnamefont {P.}~\bibnamefont
  {Blaha}}, \bibinfo {author} {\bibfnamefont {K.}~\bibnamefont {Schwarz}},
  \bibinfo {author} {\bibfnamefont {G.~K.~H.}\ \bibnamefont {Madsen}}, \bibinfo
  {author} {\bibfnamefont {D.}~\bibnamefont {Kvasnicka}},\ and\ \bibinfo
  {author} {\bibfnamefont {J.}~\bibnamefont {Luitz}},\ }\href@noop {} {\emph
  {\bibinfo {title} {WIEN2k, An Augmented Plane Wave + Local Orbitals Program
  for Calculating Crystal Properties}}}\ (\bibinfo  {publisher} {Techn.
  {Universit\"at} Wien, Austria},\ \bibinfo {year} {2001})\BibitemShut
  {NoStop}%
\bibitem [{\citenamefont {Mostofi}\ \emph {et~al.}(2008)\citenamefont
  {Mostofi}, \citenamefont {Yates}, \citenamefont {Lee}, \citenamefont {Souza},
  \citenamefont {Vanderbilt},\ and\ \citenamefont {Marzari}}]{mostofi2008}%
  \BibitemOpen
  \bibfield  {author} {\bibinfo {author} {\bibfnamefont {A.~A.}\ \bibnamefont
  {Mostofi}}, \bibinfo {author} {\bibfnamefont {J.~R.}\ \bibnamefont {Yates}},
  \bibinfo {author} {\bibfnamefont {Y.-S.}\ \bibnamefont {Lee}}, \bibinfo
  {author} {\bibfnamefont {I.}~\bibnamefont {Souza}}, \bibinfo {author}
  {\bibfnamefont {D.}~\bibnamefont {Vanderbilt}},\ and\ \bibinfo {author}
  {\bibfnamefont {N.}~\bibnamefont {Marzari}},\ }\href
  {\doibase10.1016/j.cpc.2007.11.016} {\bibfield  {journal} {\bibinfo
  {journal} {Comput. Phys. Commun.}\ }\textbf {\bibinfo {volume} {178}},\
  \bibinfo {pages} {685--699} (\bibinfo {year} {2008})},\ \Eprint
  {http://arxiv.org/abs/0708.0650} {arXiv:0708.0650
  [cond-mat.mtrl-sci]}\BibitemShut {NoStop}%
\bibitem [{\citenamefont {{Kune\v s}}\ \emph {et~al.}(2010)\citenamefont
  {{Kune\v s}}, \citenamefont {Arita}, \citenamefont {Wissgott}, \citenamefont
  {Toschi}, \citenamefont {Ikeda},\ and\ \citenamefont {Held}}]{kunes2010}%
  \BibitemOpen
  \bibfield  {author} {\bibinfo {author} {\bibfnamefont {J.}~\bibnamefont
  {{Kune\v s}}}, \bibinfo {author} {\bibfnamefont {R.}~\bibnamefont {Arita}},
  \bibinfo {author} {\bibfnamefont {P.}~\bibnamefont {Wissgott}}, \bibinfo
  {author} {\bibfnamefont {A.}~\bibnamefont {Toschi}}, \bibinfo {author}
  {\bibfnamefont {H.}~\bibnamefont {Ikeda}},\ and\ \bibinfo {author}
  {\bibfnamefont {K.}~\bibnamefont {Held}},\ }\href
  {\doibase10.1016/j.cpc.2010.08.005} {\bibfield  {journal} {\bibinfo
  {journal} {Comput. Phys. Commun.}\ }\textbf {\bibinfo {volume} {181}},\
  \bibinfo {pages} {1888--1895} (\bibinfo {year} {2010})},\ \Eprint
  {http://arxiv.org/abs/1004.3934} {arXiv:1004.3934
  [cond-mat.mtrl-sci]}\BibitemShut {NoStop}%
\bibitem [{\citenamefont {Souza}\ \emph {et~al.}(2001)\citenamefont {Souza},
  \citenamefont {Marzari},\ and\ \citenamefont {Vanderbilt}}]{souza2001}%
  \BibitemOpen
  \bibfield  {author} {\bibinfo {author} {\bibfnamefont {I.}~\bibnamefont
  {Souza}}, \bibinfo {author} {\bibfnamefont {N.}~\bibnamefont {Marzari}},\
  and\ \bibinfo {author} {\bibfnamefont {D.}~\bibnamefont {Vanderbilt}},\
  }\href {\doibase10.1103/PhysRevB.65.035109} {\bibfield  {journal} {\bibinfo
  {journal} {Phys. Rev. B}\ }\textbf {\bibinfo {volume} {65}},\ \bibinfo
  {pages} {035109} (\bibinfo {year} {2001})},\ \Eprint
  {http://arxiv.org/abs/cond-mat/0108084} {arXiv:cond-mat/0108084
  [cond-mat.mtrl-sci]}\BibitemShut {NoStop}%
\bibitem [{\citenamefont {Lichtenstein}\ and\ \citenamefont
  {Katsnelson}(1998)}]{lichtenstein1998}%
  \BibitemOpen
  \bibfield  {author} {\bibinfo {author} {\bibfnamefont {A.~I.}\ \bibnamefont
  {Lichtenstein}}\ and\ \bibinfo {author} {\bibfnamefont {M.~I.}\ \bibnamefont
  {Katsnelson}},\ }\href {\doibase10.1103/PhysRevB.57.6884} {\bibfield
  {journal} {\bibinfo  {journal} {Phys. Rev. B}\ }\textbf {\bibinfo {volume}
  {57}},\ \bibinfo {pages} {6884--6895} (\bibinfo {year} {1998})},\ \Eprint
  {http://arxiv.org/abs/cond-mat/9707127} {arXiv:cond-mat/9707127
  [cond-mat.str-el]}\BibitemShut {NoStop}%
\bibitem [{\citenamefont {Leb\`{e}gue}\ \emph {et~al.}(2006)\citenamefont
  {Leb\`{e}gue}, \citenamefont {Svane}, \citenamefont {Katsnelson},
  \citenamefont {Lichtenstein},\ and\ \citenamefont {Eriksson}}]{lebegue2006}%
  \BibitemOpen
  \bibfield  {author} {\bibinfo {author} {\bibfnamefont {S.}~\bibnamefont
  {Leb\`{e}gue}}, \bibinfo {author} {\bibfnamefont {A.}~\bibnamefont {Svane}},
  \bibinfo {author} {\bibfnamefont {M.~I.}\ \bibnamefont {Katsnelson}},
  \bibinfo {author} {\bibfnamefont {A.~I.}\ \bibnamefont {Lichtenstein}},\ and\
  \bibinfo {author} {\bibfnamefont {O.}~\bibnamefont {Eriksson}},\ }\href
  {\doibase10.1103/PhysRevB.74.045114} {\bibfield  {journal} {\bibinfo
  {journal} {Phys. Rev. B}\ }\textbf {\bibinfo {volume} {74}},\ \bibinfo {eid}
  {045114} (\bibinfo {year} {2006})}\BibitemShut {NoStop}%
\bibitem [{\citenamefont {Lang}\ \emph {et~al.}(1981)\citenamefont {Lang},
  \citenamefont {Baer},\ and\ \citenamefont {Cox}}]{lang1981}%
  \BibitemOpen
  \bibfield  {author} {\bibinfo {author} {\bibfnamefont {J.~K.}\ \bibnamefont
  {Lang}}, \bibinfo {author} {\bibfnamefont {Y.}~\bibnamefont {Baer}},\ and\
  \bibinfo {author} {\bibfnamefont {P.~A.}\ \bibnamefont {Cox}},\ }\href
  {\doibase10.1088/0305-4608/11/1/015} {\bibfield  {journal} {\bibinfo
  {journal} {J. Phys. F: Met. Phys.}\ }\textbf {\bibinfo {volume} {11}},\
  \bibinfo {pages} {121--138} (\bibinfo {year} {1981})}\BibitemShut {NoStop}%
\bibitem [{\citenamefont {Georges}\ \emph {et~al.}(1996)\citenamefont
  {Georges}, \citenamefont {Kotliar}, \citenamefont {Krauth},\ and\
  \citenamefont {Rozenberg}}]{georges1996}%
  \BibitemOpen
  \bibfield  {author} {\bibinfo {author} {\bibfnamefont {A.}~\bibnamefont
  {Georges}}, \bibinfo {author} {\bibfnamefont {G.}~\bibnamefont {Kotliar}},
  \bibinfo {author} {\bibfnamefont {W.}~\bibnamefont {Krauth}},\ and\ \bibinfo
  {author} {\bibfnamefont {M.~J.}\ \bibnamefont {Rozenberg}},\ }\href
  {\doibase10.1103/RevModPhys.68.13} {\bibfield  {journal} {\bibinfo  {journal}
  {Rev. Mod. Phys.}\ }\textbf {\bibinfo {volume} {68}},\ \bibinfo {pages}
  {13--125} (\bibinfo {year} {1996})},\ \Eprint
  {http://arxiv.org/abs/cond-mat/9510091} {arXiv:cond-mat/9510091}\BibitemShut
  {NoStop}%
\bibitem [{\citenamefont {Liebsch}\ and\ \citenamefont
  {Ishida}(2012)}]{liebsch2012}%
  \BibitemOpen
  \bibfield  {author} {\bibinfo {author} {\bibfnamefont {A.}~\bibnamefont
  {Liebsch}}\ and\ \bibinfo {author} {\bibfnamefont {H.}~\bibnamefont
  {Ishida}},\ }\href {\doibase10.1088/0953-8984/24/5/053201} {\bibfield
  {journal} {\bibinfo  {journal} {J. Phys.: Condens. Matter}\ }\textbf
  {\bibinfo {volume} {24}},\ \bibinfo {pages} {053201} (\bibinfo {year}
  {2012})},\ \Eprint {http://arxiv.org/abs/1109.0158} {arXiv:1109.0158
  [cond-mat.str-el]}\BibitemShut {NoStop}%
\bibitem [{\citenamefont {Caffarel}\ and\ \citenamefont
  {Krauth}(1994)}]{caffarel1994}%
  \BibitemOpen
  \bibfield  {author} {\bibinfo {author} {\bibfnamefont {M.}~\bibnamefont
  {Caffarel}}\ and\ \bibinfo {author} {\bibfnamefont {W.}~\bibnamefont
  {Krauth}},\ }\href {\doibase10.1103/PhysRevLett.72.1545} {\bibfield
  {journal} {\bibinfo  {journal} {Phys. Rev. Lett.}\ }\textbf {\bibinfo
  {volume} {72}},\ \bibinfo {pages} {1545--1548} (\bibinfo {year}
  {1994})}\BibitemShut {NoStop}%
\bibitem [{\citenamefont {{Koloren\v c}}\ \emph {et~al.}(2015)\citenamefont
  {{Koloren\v c}}, \citenamefont {Shick},\ and\ \citenamefont
  {Lichtenstein}}]{kolorenc2015b}%
  \BibitemOpen
  \bibfield  {author} {\bibinfo {author} {\bibfnamefont {J.}~\bibnamefont
  {{Koloren\v c}}}, \bibinfo {author} {\bibfnamefont {A.~B.}\ \bibnamefont
  {Shick}},\ and\ \bibinfo {author} {\bibfnamefont {A.~I.}\ \bibnamefont
  {Lichtenstein}},\ }\href {\doibase10.1103/PhysRevB.92.085125} {\bibfield
  {journal} {\bibinfo  {journal} {Phys. Rev. B}\ }\textbf {\bibinfo {volume}
  {92}},\ \bibinfo {pages} {085125} (\bibinfo {year} {2015})},\ \Eprint
  {http://arxiv.org/abs/1504.07979} {arXiv:1504.07979
  [cond-mat.str-el]}\BibitemShut {NoStop}%
\bibitem [{\citenamefont {Lehoucq}\ \emph {et~al.}(1998)\citenamefont
  {Lehoucq}, \citenamefont {Sorensen},\ and\ \citenamefont {Yang}}]{arpack}%
  \BibitemOpen
  \bibfield  {author} {\bibinfo {author} {\bibfnamefont {R.~B.}\ \bibnamefont
  {Lehoucq}}, \bibinfo {author} {\bibfnamefont {D.~C.}\ \bibnamefont
  {Sorensen}},\ and\ \bibinfo {author} {\bibfnamefont {C.}~\bibnamefont
  {Yang}},\ }\href {\doibase10.1137/1.9780898719628} {\emph {\bibinfo {title}
  {ARPACK Users' Guide, Solution of Large-Scale Eigenvalue Problems with
  Implicitly Restarted Arnoldi Methods}}}\ (\bibinfo  {publisher} {SIAM,
  Philadelphia, PA},\ \bibinfo {year} {1998})\BibitemShut {NoStop}%
\bibitem [{\citenamefont {Meyer}\ and\ \citenamefont {Pal}(1989)}]{meyer1989}%
  \BibitemOpen
  \bibfield  {author} {\bibinfo {author} {\bibfnamefont {H.-D.}\ \bibnamefont
  {Meyer}}\ and\ \bibinfo {author} {\bibfnamefont {S.}~\bibnamefont {Pal}},\
  }\href {\doibase10.1063/1.457438} {\bibfield  {journal} {\bibinfo  {journal}
  {J. Chem. Phys.}\ }\textbf {\bibinfo {volume} {91}},\ \bibinfo {pages}
  {6195--6204} (\bibinfo {year} {1989})}\BibitemShut {NoStop}%
\bibitem [{\citenamefont {Sakurai}(1967)}]{sakurai_aqm}%
  \BibitemOpen
  \bibfield  {author} {\bibinfo {author} {\bibfnamefont {J.~W.}\ \bibnamefont
  {Sakurai}},\ }\href@noop {} {\emph {\bibinfo {title} {Advanced Quantum
  Mechanics}}}\ (\bibinfo  {publisher} {Addison-Wesley},\ \bibinfo {year}
  {1967})\BibitemShut {NoStop}%
\bibitem [{\citenamefont {Krause}\ and\ \citenamefont
  {Oliver}(1979)}]{krause1979}%
  \BibitemOpen
  \bibfield  {author} {\bibinfo {author} {\bibfnamefont {M.~O.}\ \bibnamefont
  {Krause}}\ and\ \bibinfo {author} {\bibfnamefont {J.~H.}\ \bibnamefont
  {Oliver}},\ }\href {\doibase10.1063/1.555595} {\bibfield  {journal} {\bibinfo
   {journal} {J. Phys. Chem. Ref. Data}\ }\textbf {\bibinfo {volume} {8}},\
  \bibinfo {pages} {329--338} (\bibinfo {year} {1979})}\BibitemShut {NoStop}%
\bibitem [{\citenamefont {McMahan}\ \emph {et~al.}(1998)\citenamefont
  {McMahan}, \citenamefont {Huscroft}, \citenamefont {Scalettar},\ and\
  \citenamefont {Pollock}}]{mcmahan1998}%
  \BibitemOpen
  \bibfield  {author} {\bibinfo {author} {\bibfnamefont {A.}~\bibnamefont
  {McMahan}}, \bibinfo {author} {\bibfnamefont {C.}~\bibnamefont {Huscroft}},
  \bibinfo {author} {\bibfnamefont {R.}~\bibnamefont {Scalettar}},\ and\
  \bibinfo {author} {\bibfnamefont {E.}~\bibnamefont {Pollock}},\ }\href
  {\doibase10.1023/A:1008698422183} {\bibfield  {journal} {\bibinfo  {journal}
  {J. Comput.-Aided Mater. Des.}\ }\textbf {\bibinfo {volume} {5}},\ \bibinfo
  {pages} {131--162} (\bibinfo {year} {1998})},\ \Eprint
  {http://arxiv.org/abs/cond-mat/9805064} {arXiv:cond-mat/9805064
  [cond-mat.str-el]}\BibitemShut {NoStop}%
\bibitem [{\citenamefont {Gull}\ \emph {et~al.}(2011)\citenamefont {Gull},
  \citenamefont {Millis}, \citenamefont {Lichtenstein}, \citenamefont
  {Rubtsov}, \citenamefont {Troyer},\ and\ \citenamefont {Werner}}]{gull2011}%
  \BibitemOpen
  \bibfield  {author} {\bibinfo {author} {\bibfnamefont {E.}~\bibnamefont
  {Gull}}, \bibinfo {author} {\bibfnamefont {A.~J.}\ \bibnamefont {Millis}},
  \bibinfo {author} {\bibfnamefont {A.~I.}\ \bibnamefont {Lichtenstein}},
  \bibinfo {author} {\bibfnamefont {A.~N.}\ \bibnamefont {Rubtsov}}, \bibinfo
  {author} {\bibfnamefont {M.}~\bibnamefont {Troyer}},\ and\ \bibinfo {author}
  {\bibfnamefont {P.}~\bibnamefont {Werner}},\ }\href
  {\doibase10.1103/RevModPhys.83.349} {\bibfield  {journal} {\bibinfo
  {journal} {Rev. Mod. Phys.}\ }\textbf {\bibinfo {volume} {83}},\ \bibinfo
  {pages} {349--404} (\bibinfo {year} {2011})},\ \Eprint
  {http://arxiv.org/abs/1012.4474} {arXiv:1012.4474
  [cond-mat.str-el]}\BibitemShut {NoStop}%
\end{thebibliography}%

\end{document}